\documentclass[twocolumn,prl,showpacs,preprintnumbers,amsmath,amssymb]{revtex4}

\usepackage{graphicx}
\usepackage{dcolumn}
\usepackage{bm}
\begin{document}


\title{Comment on the "Second dip as a signature of ultrahigh energy proton interactions with cosmic microwave background radiation" by V.~Berezinsky, A.~Gazizov and K.~Kachelrie{\ss} }

\author{Tadeusz Wibig}
 \email{wibig@zpk.u.lodz.pl}
\affiliation{%
Theoretical Physics Dept., University of Lodz;\\
Cosmic Ray Lab., Soltan Institute for Nuclear Studies, Uniwersytecka 5,
90-950 Lodz, Poland.}

\date{\today}

\begin{abstract}
A new feature in the spectrum of ultra high energy cosmic rays (UHECR) has been
announced in the paper by Berezinsky, Gazizov and Kachelrie{\ss}. The ratio of the
solution of the exact transport equation to its solution in the 
continuous energy loss limit shows intriguing features which, according to the Authors,
 are related to the very nature of the energy loss processes of UHECR:
the very sharp {\it second dip} predicted
 at $6.3 \times 10^{19}$eV can be used as an energy calibration point and also as the UHECR mass indicator for
big future cosmic ray experiments. In the present paper we would like to advocate that this
statement is 
an overinterpretation. The {\it second dip} is a result of an
inapproriate approximation used, and thus it can't help to understand 
the nature of UHECR in any way.

\end{abstract}

\pacs{98.70.Sa, 13.85.Tp}
\maketitle

{\it Introduction} --- 
The nature and origin of Ultra-High Energy Cosmic Rays (UHECR) (particles above about $10^{18}$eV) has been
a mystery for a long time and it is still one of most intriguing questions of contemporary physics. It deals with the problems of the Universe as a whole, its structure and evolution, but also with the elementary components of matter, the theory of interactions at extremely high energies and it could be helpful to answer some fundamental questions about the structure of space and time like, such as the possible violation of Lorentz symmetry.

General isotropy and lack of clear point sources suggests that the UHECR
are extragalactic. If they are protons, it is known that their flux should have
a severe reduction below 10$^{20}$eV because of the cosmic microwave background radiation.
This mechanism is well-known as Greisen-Zatsepin-Kuzmin (GZK) effect. It is caused by the resonant production of $\Delta$s in interactions with the universal microwave background photons.
The UHECR could contain also heavier atomic nuclei, and they will disappear from the
extragalactic flux due to giant dipole resonance excitation. The mechanism leads to separation
of the few nucleons from the CR nucleus, diminishing its total energy by the respective fraction. This process starts to contribute at higher energy than the resonant proton energy losses, but from the point of view of the particular problem of UHECR propagation (and the {\it second dip}) both work in the same way: they are not continuous energy loss processes, but rather discrete, probabilistic processes. At any moment
there is a non-zero probability of losing a significant fraction of the particle energy. They are also, and this is important here, threshold reactions: if there are effectively no photons of the energy needed to create a respective resonance nothing can happen and UHECR will not lose energy this way.

For charged UHECR there is an additional process leading to energy dissipation. It is the creation of $e^+\:e^-$ pairs with interactions with the same microwave photons. Due
to the small mass of the electron in comparison to that of the pion or nucleon (hereafter, we'll consider protons only - in the case of nuclei the results are similar) the energy
lost by a proton in $e^+\:e^-$ production is very small and the process can be treated in a Continuous Energy Loss (CEL) way.

\

{\it UHECR transport and the second dip} ---
The description of the propagation of UHECR particles with energy losses as described above can be made with the help of the
transport equation given and described briefly in Ref.\cite{Berezinsky:2006mk}
{\it the kinetic equation}. In the present use it can be written as:
\begin{eqnarray}\label{kin}
{{\partial n(E,t)} \over {\partial t}}~=~
{\partial \over \partial E} \left[b_{\rm pair}(E) n(E,t)\right]
-P(E) n(E,t) \nonumber \\
~~~~~~+ \int\limits_E^{E_{\rm max}} dE' P(E',\:E) n(E',t) +Q(E,t)~,
\end{eqnarray}
where $b$ describes the continuous energy losses and $P$ the discrete process. $Q$ is the
source term.
In CEL limit, all the losses are included in the $b_{\rm tot}$ factor and the solution
is then
\begin{equation}\label{cel}
 n(E,t_0)\:d E~=~ \int\limits_{t_{\rm start}}^{t_0} dt Q(E_g(E)) d E_g~,
\end{equation}
where $E_g$ is the energy at the origin of the particle observed after time $t$
with the energy $E$. The relation between $E$ and $E_g$ is uniquely determined by the $b_{\rm tot}$ function.

If the GZK energy losses are approximated by the 'average GZK energy losses' and combined with the $e^+\:e^-$ production losses, the solution
of the transport equation is called (after \cite{Berezinsky:2006mk}) a {\it universal spectrum}.

The idea of the {\it second dip} 
is that the measured UHECR flux as well as that calculated, represented by the solution of Eq.(\ref{kin}),
(when divided by the {\it universal spectrum}) has a tiny but significant dip just 
about the energy where the continuous and discrete energy loss processes are equally important - at the end of the so-called 'ankle'
structure in UHECR spectrum ($6.1\times 10^{19}$ eV).

The importance of this {\it second dip} is that its position 
could be used as an energy calibration point, or to distinguish UHECR being mostly protons
(GZK losses) or nuclei (fragmentation by the Giant Dipole Resonance). 
This is one of the most important questions concerning the origin extremely high energy cosmic rays.

However, if one looks closer to the mechanism of {\it second dip} formation
it can be supposed that it is produced by the approximation used, ie it is an atrifact. 
The validity of the CEL approximation is discussed in Ref.\cite{Berezinsky:2006mk}. The linearity of the (average) energy loss rate ($b_{\rm tot}$) with respect to particle energy is one of the requirements. Another needs the smooth behaviour of the injection spectrum (the vanishing of ${{\partial \ln(n)} / {\partial \ln (E)}}$ for the UHECR transport equation which scales with $E'/E$).
It can be expected that when the energy losses rate changes
significantly (as it is in the case of UHECR, when the GZK process comes in very rapidly, almost exponentially, as mentioned in Ref.\cite{Berezinsky:2006mk}), or when the source spectrum ends (the effectiveness of the acceleration or UHECR production mechanisms, in general, has to have its limits) the
results obtained in the {\it kinematic} or CEL approximations
are different. The second point is discussed in \cite{Berezinsky:2006mk}.

\

{\it Toy model for particle transport energy losses} ---
To study the {\it second dip} creation in detail we  present results obtained with a 
simple 'toy model'. 

\begin{figure}
\includegraphics[width=7.5cm]{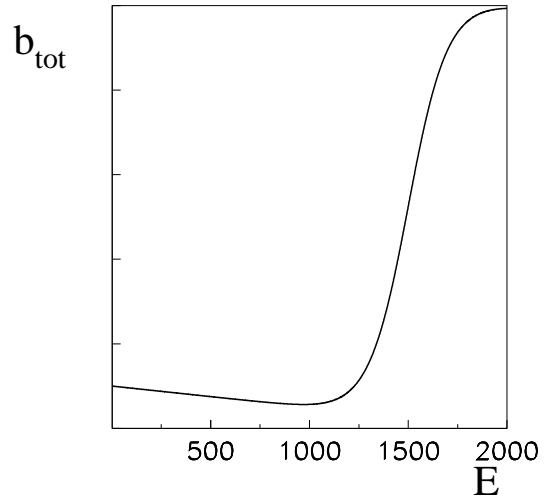}
\caption{\label{sigma} Average energy losses in our toy model.}
\end{figure}

We assumed that 
'particles', whose 'energy' spectrum (measured in some arbitrary units - a.u.)  is 
the subject of the study,
lose energy with the rate shown in Fig.\ref{sigma}, which has a profound
increase (similar, but smaller that in the UHECR case). These losses can be 
treated in two ways: once as a continuous process and secondly as a discrete
process with cross-section proportional to the average and the
distribution of particular losses given by half of a Gaussian
with the standard deviation of 2 a.u. 

\begin{figure}
\includegraphics[width=7.5cm]{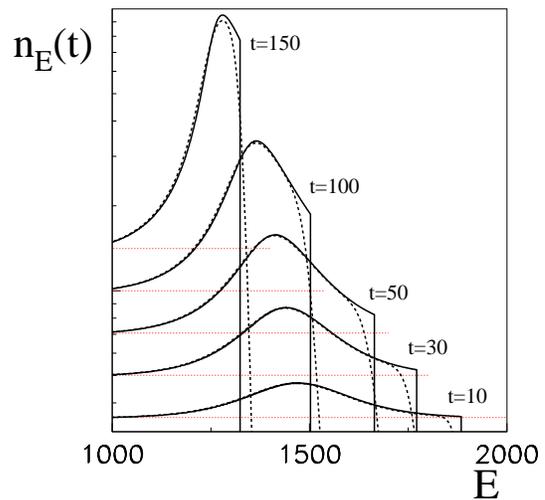}
\caption{\label{2spectra} Spectra observed after a given 'time' from
the injection instant (displaced vertically as shown by thin dotted horizontal lines). 
The solid curves show results in the CEL limit while the dashed curves are for discrete losses
description. The units of $t$ are arbitrary.}
\end{figure}

We choose an injection energy spectrum which is uniform in the energy range of interest (up to 2000 a.u.) with no particles created with energies above 2000 a.u.

Then we perform propagation calculations due to the continuous and discrete 
energy losses description. The results are as given in Fig.~\ref{2spectra}.

The details of the model such as the particular shape of $b$, the probability distribution $P$, 
and the initial spectrum $Q$ are not crucial and the results given below remain
similar also for other choices. 

First, we see that for any particular 'time' the CEL solution 
has a sharp high energy cut-off and the effect of its softening when 
the probabilistic treatment is applied is very clear. 

\begin{figure}
\includegraphics[width=7.5cm]{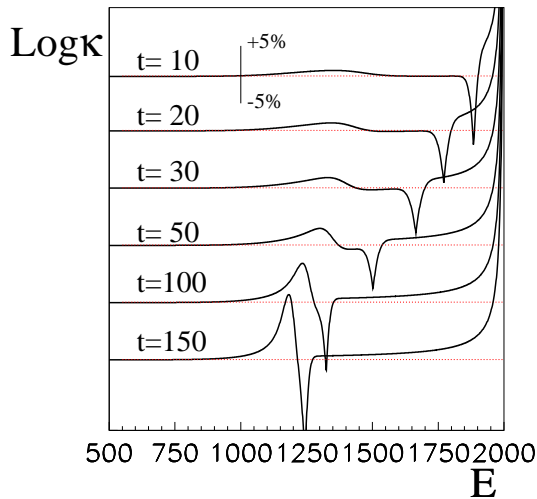}
\caption{\label{kapp1}Logarithm of the {\it distortion factor} (thick lines) for 
our toy model for different times since injection. 
Curves show the ratio of the discrete energy losses solution to 
the CEL approximation result for the source active for a given 
'time' before the present. Results for different 'times' are displaced
vertically by a factor of 10\%.
The small vertical line shown the $\pm 5\%$ change range.}
\end{figure}

The second quite obvious effect is a grouping of the 'particles' at energies 
where the losses become slower. The shape of this bump is determined mainly 
by the energy losses rate (Fig.~\ref{sigma}), but it is also slightly different for both calculation
methods. This is much better seen in Fig.~\ref{kapp1}, where we plot the 
{\it distortion factor}: the ratio of the respective spectra. 
These final spectra are obtained by integrating the particle fluxes observed 
at every moment since the injection (shown in Fig.~\ref{2spectra}).

\

{\it Results on the distortion factor $\kappa$} ---
We see  three important features in the Fig.\ref{kapp1}.
Starting from the highest energies:
\begin{itemize}
\item[{\it i} -] the obvious overabundance of the probabilistic spectra with respect to the CEL solution. This is an obvious effect of smoothing the high energy cut-off edge.
This is described clearly in Ref.\cite{Berezinsky:2006mk},

\item[{\it ii} -] a very sharp narrow dip observed for any 'time'. 
The dip moves as the propagation time increases exactly 
as the energy of the highest energy 
of the source initial spectrum (2000 a.u. in our toy model) 
propagated by the CEL 
mechanism,

\item[{\it iii} -] the bump seen at the point where the energy 
losses becomes very small. This
is an effect of the accumulation of particles which have lost their energy, 
according to the fluctuations in probabilistic treatment, much 
faster than average. This bump also moves to lower energies.
\end{itemize}

It is highly probable that the effect called in Ref.\cite{Berezinsky:2006mk}
the {\it second dip} is the one listed above as item {\it ii}. 

If the time of activity of the source were infinite, the intergration would need to
 be done up to infinity. Then both: the dip and bump disappear from the energy 
interval under study and only one remaining feature will be the smooth
 overabundance just below 2000.
But for the real case of UHECR the integration could not be longer than 
e.g. the age of the Universe. When looking at the UHECR energy losses, and 
comparing with the age of plausible UHECR sources it has to be said that it
 is quite reasonable
to put the upper limit of the integration at (about) the moment when the sharp dip
seen in Fig.\ref{kapp1} reaches the beginning of the fast rise of the energy losses.

For the real UHECR the energy losses for the GZK process are much faster
than for the $e^+\:e^-$ production, so particles produced with maximum 
energy at the sources reach the point of the beginning of GZK losses very 
quickly and then remain there for long. This leaves very small room to move 
the position of the {\it second dip} on the energy scale by changing the 
time of propagation of UHECR within reasonable limits. 

The position of the {\it second dip} doesn't depend also on the upper energy 
limit used for calculations, because of the mentioned very high rate of the 
energy losses for very high energies. It is no matter how (reasonably) big 
the upper production energy limit is, it goes rapidly to the point where the 
GZK process starts.

Additional confirmation of such an interpretation is the substantial increase in the
 {\it distortion factor} above the dip (our item {\it i}), just as we see 
in our toy model, and what has to be present in any calculations with a
sharp initial spectrum cut-off.

There is an intriguing absence of the bump (our item {\it iii}) 
in the analysis presented in Ref.~\cite{Berezinsky:2006mk}.
As is seen in Fig.\ref{kapp1} for the situation when the 'times' 
are not very big the
bump is not well formed. It is possible, in principle, to adjust 
the propagation time and the maximum energy to make it less 
pronounced, while the {\it second dip} is still visible with 
a similar shape and strength.
The second dip position reported in Ref.\cite{Berezinsky:2006mk} 
is $6.3 \times 10^{19}$eV while the balance between GZK and 
$e^+\:e^-$ energy losses is at
the energy of $6.1 \times 10^{19}$eV which can support the 
above explanation.

\

{\it Conclusion} ---
It is shown that the {\it second dip} in UHECR spectrum introduced in 
Ref.~\cite{Berezinsky:2006mk}
arises from  applying an inappropriate approximation in the reference 
{\it universal spectrum}. The combination of the abrupt end of the 
production spectrum and the substantial rise of the energy loss rate 
when the GZK process starts gives the
solution of the CEL approximation an unphysical sharp jump at some point 
(which moves on the energy scale with the time of particle propagation). The 
UHECR spectrum, which is
assumed to be described by the {\it kinetic equation}, has no strange 
features 
corresponding to the {\it second dip}.
The position and shape of the 
{\it second dip} seen in the {\it distortion factor} is, to some 
extent, related to the particular details of the energy losses and 
initial conditions applied to obtain the
CEL solution. They must both be adjusted to reproduce the data 
on the UHECR spectrum. 
This almost entirely determines these details (with reasonable 
assumptions about 
injection spectrum, intergalactic radiation densities etc.). 
This implies that
the {\it distortion factor} in any future giant cosmic ray 
experiment (e.g., EUSO) will have, by definition,  the {\it second dip}. 
However, its
existence will add no new knowledge to our understanding of
the UHECR nature and origin.


\begin{thebibliography}{1}
\expandafter\ifx\csname natexlab\endcsname\relax\def\natexlab#1{#1}\fi
\expandafter\ifx\csname bibnamefont\endcsname\relax
  \def\bibnamefont#1{#1}\fi
\expandafter\ifx\csname bibfnamefont\endcsname\relax
  \def\bibfnamefont#1{#1}\fi
\expandafter\ifx\csname citenamefont\endcsname\relax
  \def\citenamefont#1{#1}\fi
\expandafter\ifx\csname url\endcsname\relax
  \def\url#1{\texttt{#1}}\fi
\expandafter\ifx\csname urlprefix\endcsname\relax\def\urlprefix{URL }\fi
\providecommand{\bibinfo}[2]{#2}
\providecommand{\eprint}[2][]{\url{#2}}

\bibitem[{\citenamefont{Berezinsky et~al.}(2006)\citenamefont{Berezinsky,
  Gazizov, and Kachelriess}}]{Berezinsky:2006mk}
\bibinfo{author}{\bibfnamefont{V.}~\bibnamefont{Berezinsky}},
  \bibinfo{author}{\bibfnamefont{A.}~\bibnamefont{Gazizov}}, \bibnamefont{and}
  \bibinfo{author}{\bibfnamefont{M.}~\bibnamefont{Kachelriess}},
  \bibinfo{journal}{Phys. Rev. Lett.} \textbf{\bibinfo{volume}{92}},
  \bibinfo{pages}{231101} (\bibinfo{year}{2006}), \eprint{astro-ph/0612247}.

\end{thebibliography}

\end{document}